\documentclass[conference]{IEEEtran}
\IEEEoverridecommandlockouts
\usepackage{cite}
\usepackage{amsmath,amssymb,amsfonts}
\usepackage{algorithmic}
\usepackage{graphicx}
\usepackage{pspicture}
\usepackage{textcomp}
\def\BibTeX{{\rm B\kern-.05em{\sc i\kern-.025em b}\kern-.08em
    T\kern-.1667em\lower.7ex\hbox{E}\kern-.125emX}}

\usepackage{amsmath,amsthm,amssymb,mathtools}
\mathtoolsset{showonlyrefs=true} 
\usepackage{physics}
\usepackage{bbm}

\mathchardef\Re="023C
\mathchardef\Im="023D
\usepackage{breqn}
\usepackage{tikz}
\usepackage{float}
\usepackage{caption}
\usepackage{subcaption}
\usepackage{gensymb}

\newcommand*\conj[1]{\overline{#1}}
\DeclareMathOperator{\E}{\mathbb{E}}

\newcommand{\rem}[1]{}

\newcommand{\bre}{\begin{equation}}
\newcommand{\ere}{\end{equation}}

\newcommand{\ee}\]
\newcommand{\bfg}{\begin{figure}[hbtp]}
\newcommand{\efg}{\end{figure}}
\newcommand{\bver}{\begin{verbatim}}
\newcommand{\ever}{\end{verbatim}}
\newcommand{\bit}{\begin{itemize}}
\newcommand{\eit}{\end{itemize}}
\newcommand{\ben}{\begin{enumerate}}
\newcommand{\een}{\end{enumerate}}

\newcommand{\btheta}{\boldsymbol\theta}

\newcommand{\given}{\: | \:}

\newcommand{\bphi}{{\mathbf \Phi}}

\newcommand{\baa}{\begin{eqnarray*}}
\newcommand{\eaa}{\end{eqnarray*}}

\newcommand{\bh}{{\bf h}}

\newcommand{\bX}{{\bf X}}

\newcommand{\bw}{{\bf w}}

\newcommand{\bx}{{\bf x}}
\newcommand{\by}{{\bf y}}

\newcommand{\cL}{{\cal L}}

\newcommand{\cH}{{\cal H}}

\newcommand{\cN}{{\mathcal{N}}}

\newcommand{\bhh}{\hat{\bf h}}

\newcommand{\defined}{\triangleq}

\def\defined{\: {\stackrel{\scriptscriptstyle \Delta}{=}} \: }

\newfont{\boldlarge}{msbm10 scaled 1100}

\newcommand{\comment}[1]{}

\newlength{\tmpbigbar}

\begin{document}

\title{Blind Channel Equalization using Variational Autoencoders}

\author{\IEEEauthorblockN{Avi Caciularu, David Burshtein}
\IEEEauthorblockA{\textit{School of Electrical Engineering, }
\textit{Tel-Aviv University, }
Tel-Aviv, 6997801 Israel \\
avi.c33@gmail.com, burstyn@eng.tau.ac.il}
}

\maketitle

\begin{abstract}
A new maximum likelihood estimation approach for blind channel equalization, using variational autoencoders (VAEs), is introduced.
Significant and consistent improvements in the error rate of the reconstructed symbols, compared to constant modulus equalizers, are demonstrated.
In fact, for the channels that were examined, the performance of the new VAE blind channel equalizer was close to the performance of a non-blind adaptive linear minimum mean square error equalizer. The new equalization method enables a significantly lower latency channel acquisition compared to the constant modulus algorithm (CMA).
The VAE uses a convolutional neural network with two layers and a very small number of free parameters.
Although the computational complexity of the new equalizer is higher compared to CMA, it is still reasonable, and the number of free parameters to estimate is small.
\end{abstract}


\section{Introduction}
Following the amazing success of deep learning methods in various tasks,
these techniques have recently been considered in some communication problems.
For example, in \cite{nachmani2016learning,nachmani2017deep,tenbrink,cammerer2017scaling} deep learning methods were considered to the problem of channel decoding,
in \cite{AutoencoderComm} the authors proposed an autoencoder as a communication system for short block codes,
and in \cite{farsad2017detection} deep learning-based detection algorithms were used when the channel model is unknown.

Our work considers transmission over a noisy intersymbol interference (ISI) channel with an unknown impulse response. Equalization methods for ISI channels using neural networks have been dealt with extensively in the literature \cite{burse2010channel}. In this paper we consider the case where the input sequence is also unknown, and 
blind channel equalization is required.
Following the blind equalization step, one can apply decision directed equalization, using the blind equalization estimation as an initial value.
Blind channel equalization is a special type of blind deconvolution where the input is constrained to lie in some known discrete constellation with known statistics.
The standard approach to tackle this problem is the constant modulus algorithm (CMA) \cite{godard1980self,treichler1983new},\cite{johnson1998blind}. Blind neural network-based algorithms using the constant modulus (CM) criterion have also been proposed in the literature \cite{you1998nonlinear}.

In this work we propose a new approach to blind channel equalization using the maximum likelihood (ML) criterion.
The ML criterion has already been used for blind channel equalization \cite{ghosh1992maximum,tong1998multichannel,cirpan1999maximum} (and references therein). However, the proposed solutions use the expectation maximization (EM) algorithm or an approximated EM, that require an iterative application of the forward-backward or Viterbi algorithms. The complexities of these algorithms are exponential in the channel memory size, which may be prohibitive. As an alternative, in this paper we propose an approximated ML estimate using the variational autoencoder (VAE) method \cite{kingma2013auto,rezende2014stochastic}. VAEs are widely used in the literature of deep learning for unsupervised and semi-supervised learning, and as a generative model to a given observations data.
We demonstrate significant and consistent improvements in the quality of the detected symbols compared to the baseline blind equalization algorithms.
In fact, for the channels that were examined, the performance of the new VAE blind channel equalizer (VAEBCE) was close to the performance of a non-blind adaptive linear minimum mean square error (MMSE) equalizer \cite{gong2010adaptive}. The new equalization method enables lower latency acquisition of an unknown channel response.
Although the computational complexity of the new VAEBCE is higher compared to CMA, it is still reasonable, and the number of free parameters to estimate is small.

\section{Problem setup}
\begin{figure}[H]
	\centering
	\includegraphics[width=.90\linewidth]{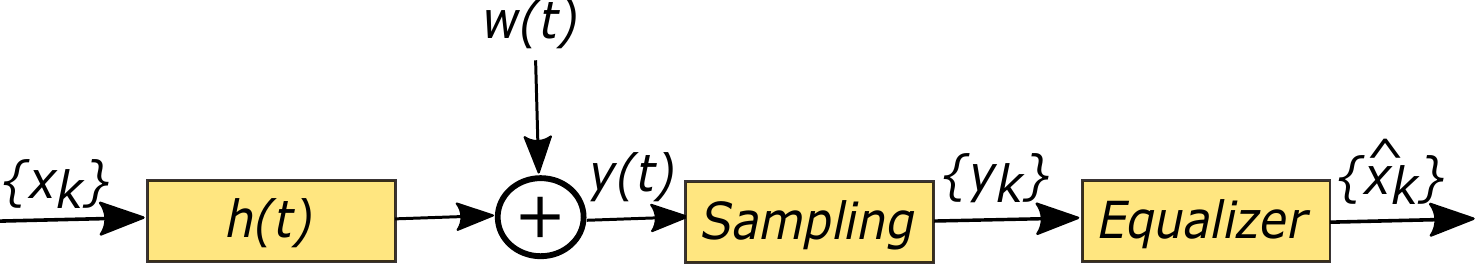}
	\caption{End to end communication system model}
	\label{fig:sys}
\end{figure}

The communication channel is modeled as a convolution of the input, $\{ x_k \}$, with some causal, finite impulse response (FIR), time invariant filter, $\bh=(h_0,h_1,\ldots,h_{M-1})$, of size $M$, followed by the addition of white Gaussian noise
\begin{equation}
	y_n=\sum_{k} x_k h_{n-k} + w_n
	\label{eq:sys}
\end{equation}
This is the equivalent model of $\{ y_k \}$ in the end to end communication system shown in Fig. \ref{fig:sys}, where the sampling is performed at the symbol rate.
The equalizer in Fig. \ref{fig:sys} reconstructs an estimate of the transmitted symbol sequence, $\{\hat{x}_k\}$. Now, suppose that we observe a finite window of measurements data $\by\defined (y_0,y_1,\ldots,y_{N-1})$.
For clarity of presentation we assume that the input signal is causal ($x_k=0$ for $k<0$). We refer to this assumption later.
Equation \eqref{eq:sys} can be written compactly for the measurements collected in $\by$ as
\begin{equation}
\by = \bx * \bh + \bw
\label{eq:conv}    
\end{equation}
where $\bx=(x_0,x_1,\ldots,x_{N-1})$ is the transmitted message, and $\bw=(w_{0},w_1,\ldots,w_{N-1})$ is an i.i.d. sequence of additive white Gaussian noise.
Throughout the paper we assume QPSK modulation, although the derivation can be extended to other constellations. 
Hence, $x_k = \pm 1 \pm j$, and the above vectors can be written as combinations of real ($I$) and imaginary ($Q$) components, so that, $\bx=\bx^{I}+j\cdot \bx^{Q}$, $\bh=\bh^{I}+j\cdot \bh^{Q}$ and $\by=\by^{I}+j\cdot \by^{Q}$.
Each element of the noise sequence, $\bw$, is complex Gaussian with statistically independent real and complex components, and with variance $\sigma_w^2$. 
Given $\bx$, $\by^I$ and $\by^Q$ are statistically independent, normally distributed.
The conditional density function of $\by^I$ is $\cN(\Re \left(\bx*\bh\right),(\sigma^{2}_{w}/2) I_N)$. The conditional density function of $\by^Q$ is $\cN(\Im \left(\bx*\bh\right),(\sigma^{2}_{w}/2) I_N)$.
Thus, for $\btheta \defined \left\{ \bh, \sigma_w^2 \right\}$, the conditional density of $\by$ given $\bx$ can be expressed as 
\begin{align}
p_{\btheta} (\by \given \bx) 
&= p_{\btheta} (\by^I \given \bx^I) p_{\btheta} (\by^Q \given \bx^Q) \nonumber\\
&=
\frac{1}{\left(\pi \sigma^{2}_{w} \right)^{N}} \cdot
e^{-\left\lVert \by-\bx*\bh \right \rVert^{2} / \sigma^{2}_{w}} 
\label{eq:pyx}
\end{align}

\section{Proposed model}
We propose using ML estimation of the channel impulse response, $\bh$. That is, we search for the vector $\bh$ and channel noise variance, $\sigma_w^2$, that maximize $\log p_\btheta(\by)$\footnote{The default base of the logarithms in this paper is $e$.}. The ML estimate has strong asymptotic optimality properties, and in particular asymptotic efficiency \cite{tong1998multichannel}. For the CMA criterion, on the other hand, one can only claim asymptotic consistency \cite{shalvi1990new}.
However, applying the accurate ML criterion to our problem is very difficult since $p_\btheta(\by)$ should first be expressed as a multi-dimensional integral
\begin{equation*}
p_\btheta(\by) = \int_{\bx} p(\bx) p_\btheta(\by \given \bx) d \bx
\end{equation*}
where we integrate over all possible input sequences $\bx$ and where
\begin{equation}
p(\bx)= p(\bx^I) p(\bx^Q) = 2^{-2N}
\label{eq:px}
\end{equation}
since we assume a uniformly distributed transmitted sequence.
However, for this kind of problem, it has been shown in various applications that it is possible to dramatically simplify the estimation problem by using the variational approach for ML estimation\cite{kingma2013auto,rezende2014stochastic}.
By the VAE approach, instead of directly maximizing $p_\btheta(\by)$ over $\btheta$, one maximizes a lower bound as follows. It can be shown \cite{kingma2013auto} that
\begin{dmath*}
\log p_{\btheta} (\by) \geq \E_{q_{\bphi}(\bx\given\by)}
\left[-\log q_{\bphi}(\bx\given\by)+\log p_{\btheta} (\bx,\by)\right]
=
\underbrace{-D_{KL} \left[ q_{\bphi}(\bx\given\by) || p(\bx)\right]}_{A}
+
\underbrace{\E_{q_{\bphi}(\bx\given\by)}
\left[\log p_{\btheta}(\by\given\bx)\right]}_{B} \defined -\cL\left(\btheta,\bphi, \by \right)
\end{dmath*}
where $D_{KL}[\cdot || \cdot]$ denotes the Kullback Leibler distance between two density functions, and $q_\bphi(\bx\given\by)$ is an arbitrary parametrized (by $\bphi$) conditional density function.
Now, instead of directly maximizing $p_\btheta(\by)$, one maximizes the lower bound $-\cL \left(\btheta,\bphi,\by \right)$ over $\btheta$ and $\bphi$.
In fact, it can be shown \cite{kingma2013auto} that by searching over $\btheta$ and \emph{all} possible conditional densities $q(\bx\given\by)$, one obtains the ML estimate of $\btheta$.
Typically, both $p_\btheta(\by\given\bx)$ and $q_\bphi(\bx\given\by)$ are implemented by neural networks.
In our problem, $p(\bx)$ is given in \eqref{eq:px}, and \emph{the encoder}, $p_\btheta(\by\given\bx)$, is given in \eqref{eq:pyx}. We use the following model for the \emph{decoder}, $q_{\bphi}(\bx\given\by)$.
\begin{equation*}
q_{\bphi}(\bx\given\by)
=
\prod_{j=0}^{N-1} q_{\bphi}(x_{j}|\by)
=
\prod_{j=0}^{N-1} q_{\bphi}(x_{j}^{I}\given\by)q_{\bphi}(x_{j}^{Q}\given\by) 
\end{equation*}
Recalling that $x_j^I \in \{-1,1\}$ and $x_j^Q \in \{-1,1\}$, this is a multivariate Bernoulli distribution with statistical independence between components. 

\begin{figure}[H]
\centering
\includegraphics[width=0.56\linewidth]{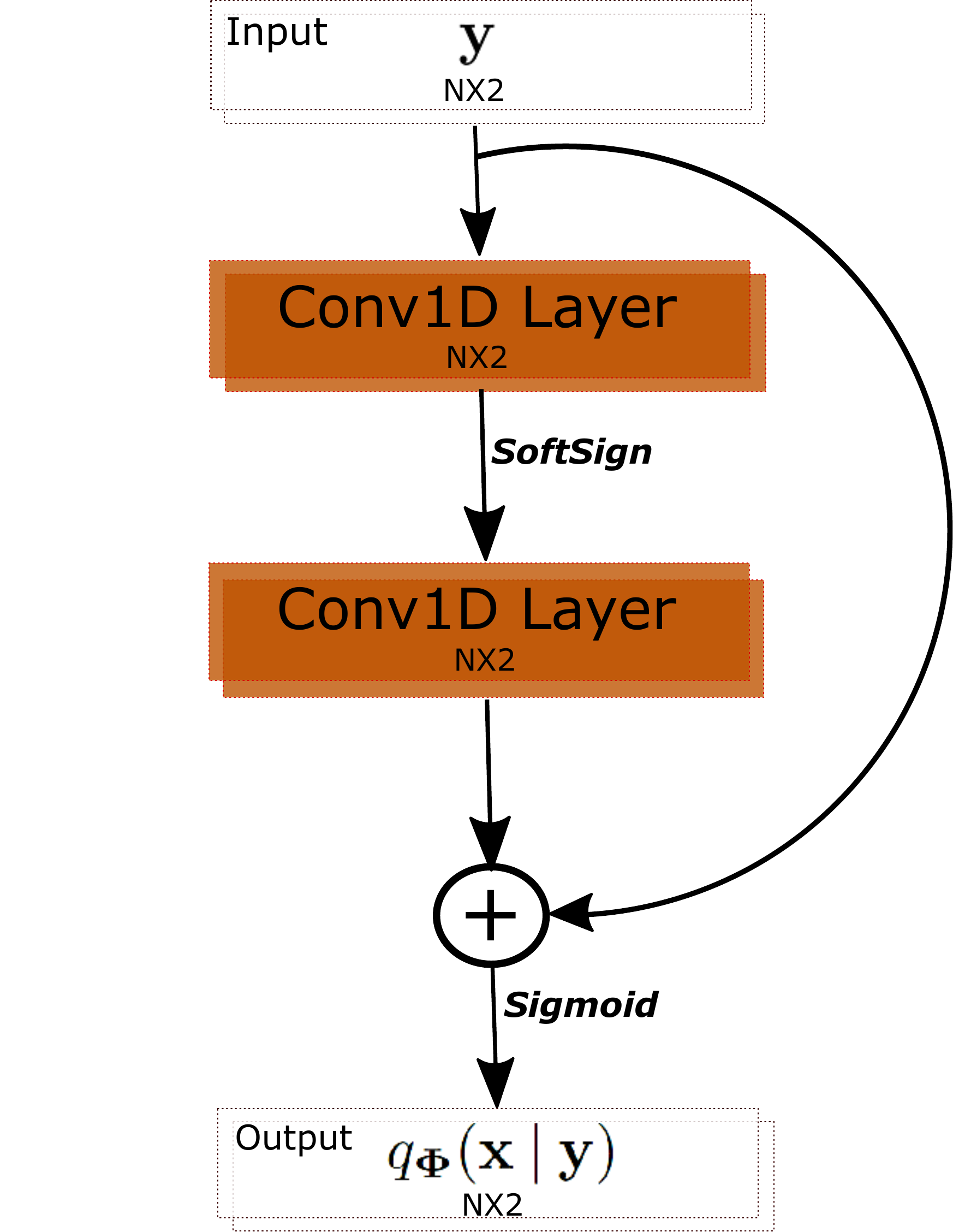}
   \caption{Our equalizer's architecture using simple fully convolutional ResNet block. Each convolution output is listed as $\text{Width}\cross \#\text{Channels}$}.
\label{fig:architecture}
\end{figure}

In our implementation of the decoder, which acts as the equalizer, we used a fully convolutional network (FCN) architecture with two convolutional layers, each with two output channels, corresponding to the real and imaginary parts of the convolution as in \cite{o2016radio}. The input and output layers are also separated to two channels corresponding to the real and imaginary components of the input, $\by$, and the output probabilities, $q$. The convolutional layers are both one dimensional (1D) as in \cite{o2016radio}, and with a residual connection as in \cite{resnet}. The non-linear activation function of the first layer is a SoftSign function defined by 
$
f\left(x\right)=\frac{x}{\abs{x}+1}
$,
which, in our experiments, proved to converge faster than other functions such as LeakyReLU and tanh. The non-linear activation function of the second layer is a sigmoid function, that ensures that the outputs are in $[0,1]$, and so they represent valid probability values. Our decoder neural network is depicted in Fig. \ref{fig:architecture}.

We now derive an explicit expression for the loss $\cL\left(\btheta,\bphi, \by \right)=-A-B$ that needs to be minimized with respect to both $\btheta$ and $\bphi$ (alternatively, $-\cL\left(\btheta,\bphi, \by \right)$ needs to be maximized).

Denote
\begin{align*}
	q_{j}^{I} &\defined q_{\bphi}(x_{j}^{I}=1|\by)\\
	q_{j}^{Q} &\defined q_{\bphi}(x_{j}^{Q}=1|\by)
\end{align*}
For the term $A$ we have
\begin{align}
A &= \sum_{\bx} q_{\bphi}(\bx|\by) \cdot \left(  \log p(\bx) - \log q_{\bphi}(\bx|\by) \right)\\
  &= \sum_{\bx} q_{\bphi}(\bx|\by) \cdot \left(  -2N\log 2 - \log q_{\bphi}(\bx|\by) \right)\\
  &= -2N\log 2 + \cH \left[ q_{\bphi}(\bx|\by) \right]
  \label{eq:A}
\end{align}
where $\cH \left[ q_{\bphi}(\bx|\by) \right]$, the entropy of $q_{\bphi}(\bx|\by)$, is given by
\begin{align}
\MoveEqLeft[4]
\mathcal{H}\left[ q_{\bphi}(\bx|\by) \right]
=
\mathcal{H}\left[ \prod_{j=0}^{N-1}q_{\bphi}(x_{j}|\by) \right]
=
\sum_{j=0}^{N-1}  \cH \left[ q_{\bphi}(x_{j}|\by) \right]
\\
\sum_{j=0}^{N-1} &\left\{
-q_{j}^{I} \log q_{j}^{I} - \left(1 - q_{j}^{I}\right) \log \left(1 - q_{j}^{I}\right)
\right. \\
&\left. -q_{j}^{Q} \log q_{j}^{Q} - \left(1 - q_{j}^{Q}\right) \log \left(1 - q_{j}^{Q}\right) \right\}
\label{eq:Hq}
\end{align}
For the term $B$ we have
\begin{align}
B&=\E_{q_{\bphi}(\bx|\by)}\left[-N\log \left(\pi \sigma_{w}^{2}\right) - \frac{1}{\sigma_{w}^{2}}\left\lVert \by-\bx*\bh \right \rVert^{2} \right]\\
&=-N\log \left( \pi \right) -N\log \left(\sigma_{w}^{2}\right)\\
&\quad - \frac{1}{\sigma_{w}^{2}}\cdot \underbrace{\E_{q_{\bphi}(\bx|\by)}\left[\left\lVert \by-\bx*\bh \right \rVert^{2} \right]}_{C}
\label{eq:B}
\end{align}
We now compute the term $C$ analytically. This is possible due to the special structure of the problem, since the generator model is linear.
This analytic computation cannot be implemented for VAEs in general. Instead, when the random variable $\bX$ in the model is continuous (e.g., a Gaussian random variable), the reparameterization trick is used \cite{kingma2013auto}. For discrete $\bX$ (as in our problem), the reparametrization trick cannot be applied. Recently, approximations for discrete $\bX$ have been proposed in \cite{maddison2016concrete}.
First, by the definition of $C$ we have,
\begin{align}
C
&=
\sum_{n=0}^{N-1}
\left[\left|y_{n}\right|^{2}-2\Re \left(y_{n}\sum_{k=0}^{N-1}\E_{q_{\bphi}(\bx|\by)}\left\{\conj{x_{k}}\right\}\conj{h_{n-k}}\right) \right.\\
&\qquad\quad
\left.
+\sum_{k,l=0}^{N-1}\E_{q_{\bphi}(\bx|\by)}\left\{x_{k}\conj{x_{l}}\right\}h_{n-k}\conj{h_{n-l}} 
\right]
\label{eq:C}
\end{align}
where $\overline{(\cdot)}$ denotes the complex conjugate. Now, 
\begin{equation}
\E_{q_{\bphi}(\bx|\by)}\left\{x_{k}\right\} = \left(2q_{k}^{I}-1\right)+j\cdot \left(2q_{k}^{Q}-1 \right)
\label{eq:Exk}
\end{equation}
Hence, for the case where $k\neq l$ we have
\begin{align}
\MoveEqLeft[0.1]
\E_{q_{\bphi}(\bx|\by)}\left\{x_{k}\conj{x_{l}}\right\} = 
\E_{q_{\bphi}(\bx|\by)}\left\{x_{k}\right\} \cdot \E_{q_{\bphi}(\bx|\by)}\left\{\conj{x_{l}}\right\}
\label{eq:Exkxl}
\\
&=\left[ \left(2q_{k}^{I}-1\right)\left(2q_{l}^{I}-1\right) + \left(2q_{k}^{Q}-1\right)\left(2q_{l}^{Q}-1\right)\right] +
\\
&\quad j\cdot 
\left[ \left(2q_{k}^{Q}-1\right)\left(2q_{l}^{I}-1\right)-\left(2q_{k}^{I}-1\right)\left(2q_{l}^{Q}-1\right)\right]
\label{eq:Exkxl1}
\end{align}
We also have
\begin{equation}
\E_{q_{\bphi}(\bx|\by)}\left\{x_{k}\conj{x_{k}}\right\} = 2
\label{eq:Exkxk}
\end{equation}
Using \eqref{eq:Exk}, \eqref{eq:Exkxl1} and \eqref{eq:Exkxk} in \eqref{eq:C}, it is straight-forward to obtain an explicit expression for $C$. However, in order to compute the third term in the summation over $n$ efficiently, we use the fact that
\begin{align}
\MoveEqLeft[2]
\sum_{k,l=0}^{N-1}\E_{q_{\bphi}(\bx|\by)}\left\{x_{k}\conj{x_{l}}\right\}h_{n-k}\conj{h_{n-l}} 
=\\
&\left| \sum_{k=0}^{N-1} \E_{q_{\bphi}(\bx|\by)}\left\{x_{k}\right\}h_{n-k} \right|^2
+\\
&\sum_{k=0}^{N-1} \left| h_{n-k} \right|^{2}
\left[2- \left| \E_{q_{\bphi}(\bx|\by)}\left\{x_{k}\right\}\right|^{2} \right]
\label{eq:C1}
\end{align}
which follows from \eqref{eq:Exkxl} and \eqref{eq:Exkxk}. It is now straightforward to use \eqref{eq:Exk}, \eqref{eq:Exkxl1}, \eqref{eq:Exkxk} and \eqref{eq:C1} in \eqref{eq:C}, and obtain
\begin{align}
C = \sum_{n=0}^{N-1} \left[ \left| y_n \right|^2 - 2\alpha_n + \beta_n \right]
\label{eq:C2}
\end{align}
where $-2\alpha_n$ and $\beta_n$, the second and third terms in the summation over $n$ in \eqref{eq:C}, are given by
\begin{align}
\alpha_n
&=
\sum_{k=0}^{N-1} \left\{ y_{n}^{I}\cdot \left[h_{n-k}^{I}\left(2q_{k}^{I}-1\right) - h_{n-k}^{Q}\left(2q_{k}^{Q}-1\right)\right] + \right.\\
&\qquad \left. y_{n}^{Q}\cdot \left[h_{n-k}^{Q}\left(2q_{k}^{I}-1\right) + h_{n-k}^{I}\left(2q_{k}^{Q}-1\right)\right] \right\}
\label{eq:alpha}
\end{align}
and
\begin{align}
\beta_n
&=
\left[\sum_{k=0}^{N-1}h_{n-k}^{I}\left(2q_{k}^{I}-1\right) - h_{n-k}^{Q}\left(2q_{k}^{Q}-1\right)\right]^{2}\\
&+
\left[\sum_{k=0}^{N-1}h_{n-k}^{Q}\left(2q_{k}^{I}-1\right) + h_{n-k}^{I}\left(2q_{k}^{Q}-1\right)\right]^{2}\\
&+
\sum_{k=0}^{N-1}\left[ \left( h_{n-k}^{I}\right)^{2}+\left( h_{n-k}^{Q}\right)^{2}\right] \cdot\\
&\qquad\quad
\left[4q_{k}^{Q} + 4q_{k}^{I}-4\left(q_{k}^{I}\right)^{2} - 4\left(2q_{k}^{Q} \right)^{2} \right]
\label{eq:beta}
\end{align}
Now, we need to minimize $-A-B$ with respect to $\btheta=\{\bh,\sigma_w^2\}$ and $\bphi$. We start with the minimization with respect to $\sigma_w^2$. Now, $A$ is independent of $\sigma_w^2$, and $B$ depends on $\sigma_w^2$ as described in \eqref{eq:B}. It is easy to see (by setting the derivative of $-B$ with respect to $\sigma_w^2$ to zero), that the optimal value of $\sigma_w^2$ is given by
$\sigma_w^2 = C / N$.
Hence, up to an additive constant (which does not influence the gradients of the learned parameters $\btheta,\bphi$), the loss function $\cL(\bh,\bphi,\by)$ (using the optimal $\sigma_w^2$) is given by
\begin{equation}
\cL(\bh,\bphi,\by) = N \log C - A
\label{eq:loss}
\end{equation}
where $A$ is given in \eqref{eq:A}-\eqref{eq:Hq}, and $C$ is given in \eqref{eq:C2}, \eqref{eq:alpha} and \eqref{eq:beta}.

We assumed that the input signal $\{x_k\}$ is causal. In reality, we are considering a block $\by$ of $N$ measurements of the signal at some arbitrary time. Therefore the above causality assumption does not hold. However, the edge effect decays as $N$ increases.
The causality assumption is equivalent to $M-1$ zero-padding of $\bx$ on the left, such that the convolution with $\bh$ according to \eqref{eq:conv} results in $\by$ of size $N$. Alternatively (supposing odd $M$ for simplicity), we assume that $\bh=(h_{-(M-1)/2},\ldots,h_0,\ldots,h_{(M-1)/2})$. Accordingly, we apply zero-padding of $\bx$ by $(M-1)/2$ both on the left and on the right. After the convolution in \eqref{eq:conv}, $\by$ is once again of size $N$. We used this second approach in our experiments, although the performance was similar to the performance of the first approach.
In all our experiments we used a mini-batch operation mode, where for each gradient descent parameters update step, we considered the gradient of the loss \eqref{eq:loss}, using only a sub-sequence of the training data, $\by$ (each update with a different sub-sequence).

Note that our loss function, $\cL(\bh,\bphi,\by)$, consists of a data entropy term, which we wish to maximize due to the i.i.d assumption of the symbols, and an autoencoder distortion term.
Also note that our method also provides an estimated channel response, as part of the learning process.

\section*{Simulation Results}
We implemented our blind equalizer using the Tensorflow framework \cite{abadi2016tensorflow} which provides automatic differentiation of the loss function.
Our algorithm was compared with the adaptive CMA \cite{abrar2010adaptive}, and with the neural network CMA (NNCMA) \cite{you1998nonlinear} blind equalization algorithms. We also compared to the linear neural network in \cite{fang1999blind}, but for clarity we did not include these results in the graphs
since the blind NNCMA outperformed the linear neural network.
In addition, we compared the performance to the adaptive MMSE \cite{gong2010adaptive} non-blind equalizer that observes the actual transmitted sequence.
The baseline algorithms use a single pass over the data for training. In order to improve performance, they were modified to have sufficiently many passes over the data.
In all our experiments, we used the Adam algorithm \cite{kingma2014adam} to minimize our loss function. The complexity of Adam is similar to that of plain gradient descent. Note, for all experiments and all blind equalization methods, that one can recover the transmitted bits only up to some unknown delay and rotation of the constellation, which for QPSK means that we need to examine four different possible rotations ($0\degree,90\degree,180\degree,270\degree)$.
The results presented in the following experiments were obtained by averaging over 20 independent training data. For each training data, we used 10,000 test data symbols to calculate the symbol error rate (SER) defined by
\begin{equation*}
\text{SER} \defined \frac{1}{M} \sum_{s}\mathbbm{1}\left( \hat{s}\neq s\right)
\end{equation*}
where $M=\text{10,000}$, $s$ is a single transmitted QPSK test symbol, $\hat{s}$ is the corresponding estimated symbol, and $\mathbbm{1}\left(\cdot\right)$ is the indicator function.

In all our experiments, we used the same FCN decoder architecture in Fig. \ref{fig:architecture}, with a filter with five complex coefficients in the first layer, and a filter with two complex coefficients in the second layer. Hence, the total number of free parameters in the model was the size of the estimated channel impulse response in the encoder in addition to only 14 ($2\times(5+2)$) real parameters in the FCN decoder.

\begin{figure}
\centering
\includegraphics[width=1.0\linewidth]{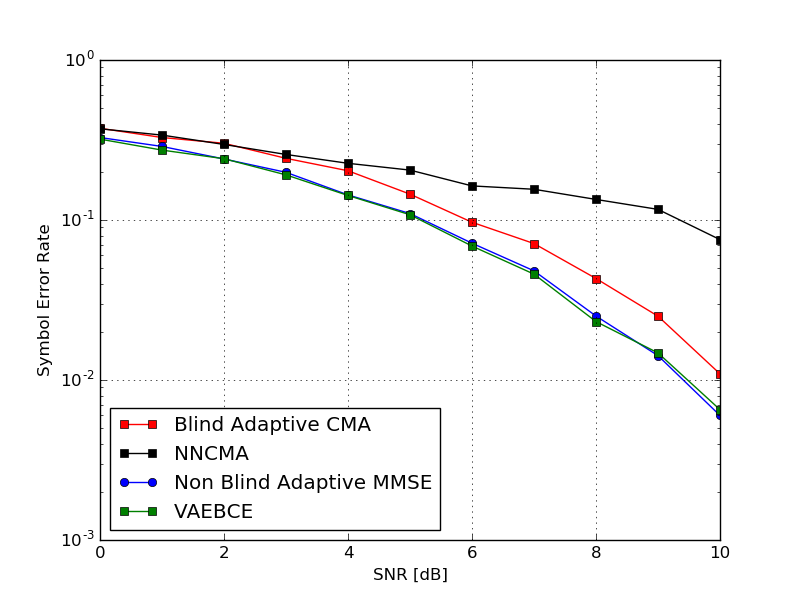}
\caption{SER vs. SNR for the equalization algorithms. The channel is $h_{1}$.}
\label{fig:snr1}
\end{figure}

\begin{figure}
\centering
\includegraphics[width=1.0\linewidth]{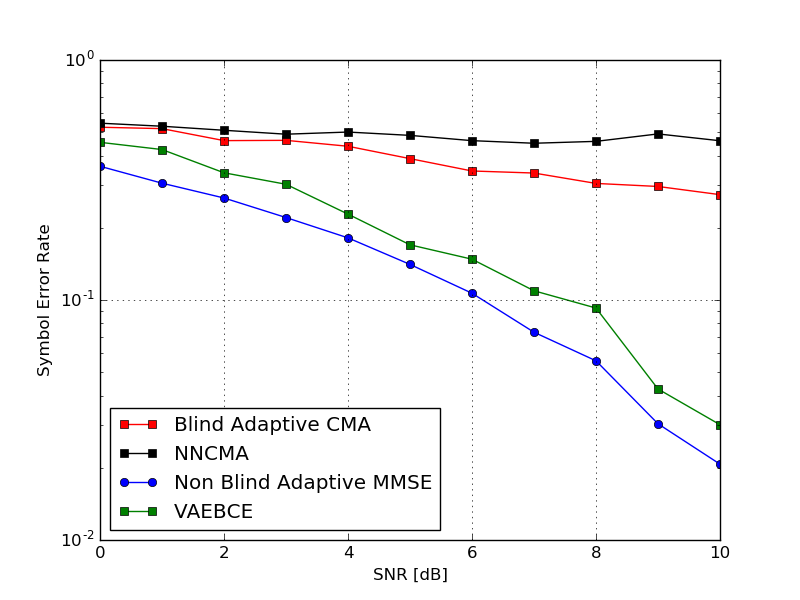}
\caption{SER vs. SNR for the equalization algorithms. The channel is $h_{2}$.}
\label{fig:snr2}
\end{figure}

\begin{figure}
\centering
\includegraphics[width=1.0\linewidth]{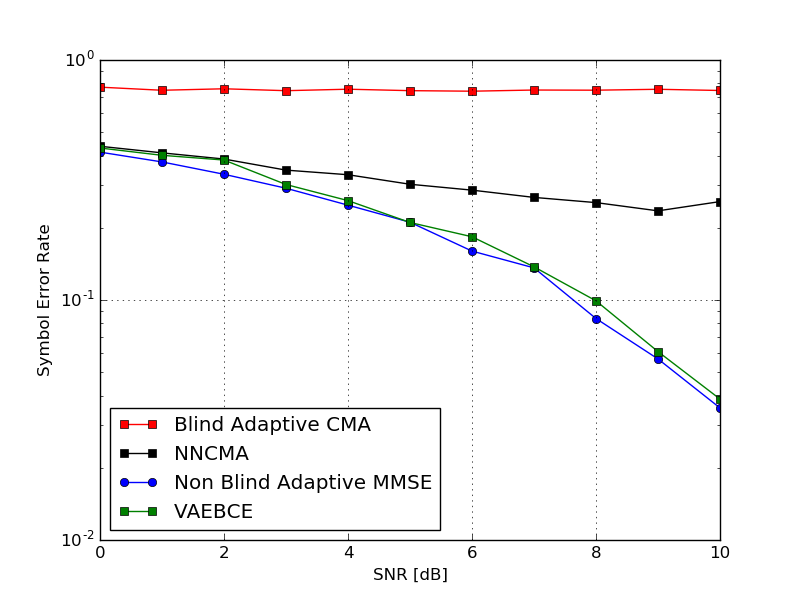}
\caption{SER vs. SNR for the equalization algorithms. The channel is $h_{3}$.}
\label{fig:snr3}
\end{figure}

In our first set of experiments,
we compared our model to the baseline algorithms at various noise levels, using the following non-minimum phase channels taken from \cite{ranhotra2017performance,fang1999blind,you1998nonlinear},
\begin{equation*}
\begin{split}
h_{1} =& [0.0545+0.05j, 0.2832-0.11971j, -0.7676+0.2788j,\\& -0.0641-0.0576j, 0.0466-0.02275j]\\
h_{2} =& [0.0554+0.0165j, -1.3449-0.4523j,\\ & 1.0067+1.1524j, 0.3476+0.3153j]\\
h_{3} =& [0.0410+0.0109j, 0.0495+0.0123j, 0.0672+0.017j,\\ & 0.0919+0.0235j,  0.7920+0.1281j, 0.396+0.0871j,\\ & 0.2715+0.048j, 0.2291+0.0415j, 0.1287+0.0154j,\\ &  0.1032+0.0119j]
\end{split}
\end{equation*}
We generated 2000 QPSK random symbols as the training set.
Then we applied convolution with the channel impulse response, and added white Gaussian noise at a signal to noise ratio (SNR) in the range $0$dB -- $10$dB. The SNR is defined by
$
\text{SNR} \defined 20\log_{10} \left( \norm{\bx*\bh} / \norm{\bw} \right)
$.
To train the model, for each update step, we sampled from the training set a mini-batch of a single sub-sequence of length $N=128$.
Figs. \ref{fig:snr1}, \ref{fig:snr2} and \ref{fig:snr3} present SER results for $h_1$, $h_2$ and $h_3$, respectively.
As can be seen, the new VAEBCE significantly outperforms the baseline blind equalizers, and is quite close to the performance of the non-blind adaptive MMSE.

\begin{figure}
\centering
\includegraphics[width=1.0\linewidth]{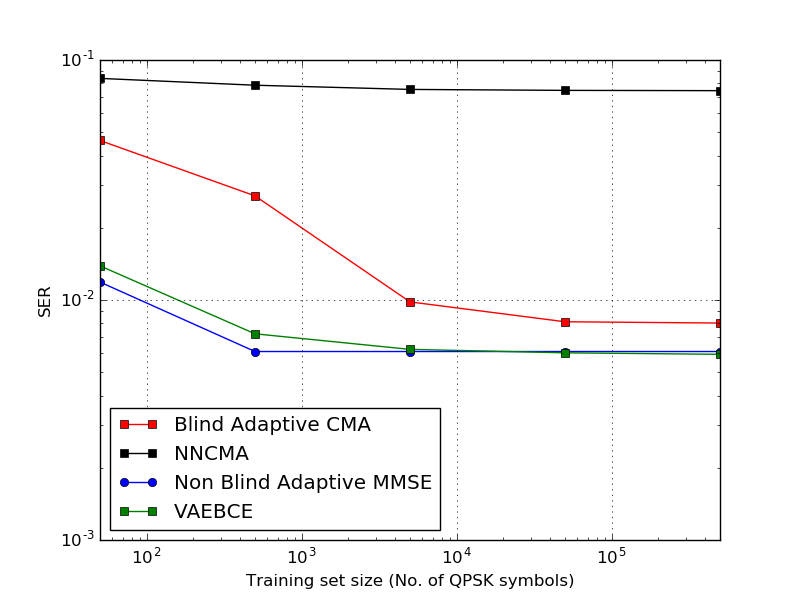}
\caption{SER vs. number of training symbols for the equalization algorithms. $\text{SNR}=10$dB. The channel used is $h_{1}$.}
\label{fig:data10}
\end{figure}

In our following experiment, we compared the SER of the equalization algorithms as the number of training symbols varied from $L=\text{50}$ to $L=\text{500,000}$. For each update step we sampled from the training set a mini-batch of a single sub-sequence of length $N=\min\left(128,L\right)$. We used the channel impulse response $h_1$ above.
Fig. \ref{fig:data10} presents the results for SNR=10dB.
Again, the new VAEBCE algorithm significantly outperforms the baseline blind equalization algorithms.

\begin{figure}
\centering
\includegraphics[width=1.0\linewidth]{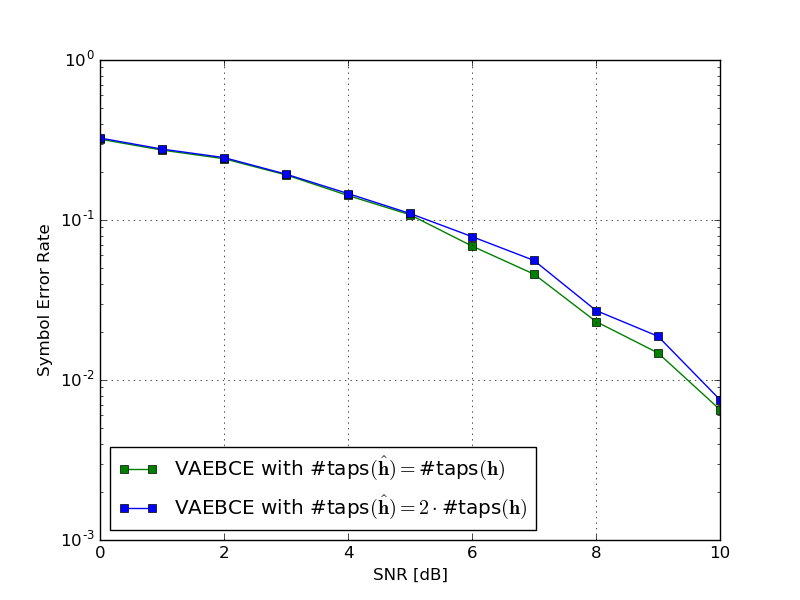}
\caption{SER vs. SNR for different lengths of $\bhh$ for the VAEBCE. The channel used is $h_{1}$.}
\label{fig:snr_diff_lengths}
\end{figure}

Recall that, in accordance with our loss function, as part of the model training we also learn an estimated channel impulse response.
We now assess the robustness to the length of the estimated channel impulse response.
Denote by $\bh$ and $\bhh$, the actual and estimated channel impulse responses, respectively.
First suppose that the length of $\bhh$ is set equal to the length of $\bh$.
In general (up to delay and rotation, as was noted above), when the SER after equalization was small, we observed a small $L_2$ distance, $||\bh-\bhh||$. This distance was monotonically decreasing with the SNR.
When the length of $\bhh$ was smaller than the length of $\bh$, the model appeared to learn $\bhh$ such that it was nearly equal to the central part of $\bh$.
When the length of $\bhh$ was larger than the length of $\bh$, the model appeared to learn an approximately zero-padded (both on the left and on the right) version of $\bh$.
In Fig. \ref{fig:snr_diff_lengths} we reevaluated the SER results when the length of $\bhh$ was twice the length of $\bh$, and did not observe a significant degradation. 

\begin{figure}
\centering
\includegraphics[width=1.0\linewidth]{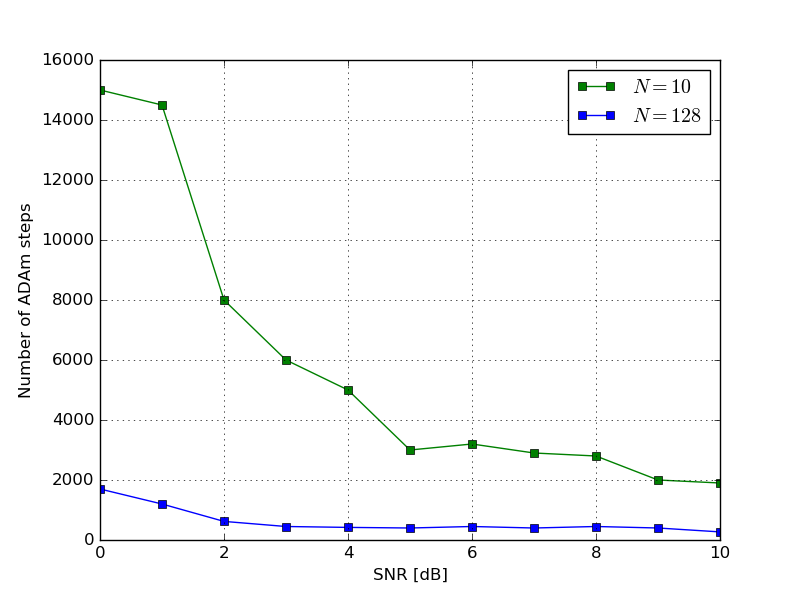}
\caption{Number of parameter updates vs. SNR for different $N$. The channel used is $h_{1}$.}
\label{fig:snr_iters}
\end{figure}

Finally, we report on the number of parameter updates required for convergence of our VAEBCE algorithm. We generated the data as described in the first experiment. To train the model, we sampled a mini-batch of a single sub-sequence of length $N\in\{10,128\}$ out of the given training symbols.
Then we let the algorithm train until convergence was achieved. The number of iterations for the channel $h_1$ is reported in Fig. \ref{fig:snr_iters}. As either $N$ or the SNR increase, the number of required iterations decreases.
For the channel $h_3$, the number of iterations was similar. Note that using the ML equalization algorithms in \cite{ghosh1992maximum,tong1998multichannel,cirpan1999maximum}, the Viterbi or forward-backward algorithms would require a trellis diagram with $2^{18}$ states, each time step. Hence, our method provides an attractive alternative.

\section*{Conclusion}
We introduced a novel algorithm for blind channel equalization using VAE (VAEBCE). We showed significantly improved SER performance compared to the baseline CMA-based blind channel equalization algorithms. In particular, VAEBCE required significantly less training symbols for the same SER measure. In fact, the performance of the new VAEBCE equalizer was close to the performance of the supervised linear adaptive MMSE equalizer.
Our equalizer is a simple FCN. This should be contrasted with alternative ML blind equalization methods, that require a trellis-based equalizer which may be much more costly to implement.
Future research should extend our method to generalized setups such as higher constellations (by replacing the output sigmoid in our FCN with a softmax).

\section*{Acknowledgment}
This research was supported by the Israel Science Foundation, grant no. 1082/13. We would like to thank Sarvraj Singh Ranhotra for sharing with us the simulations code in \cite{ranhotra2017performance}.

\bibliographystyle{IEEEtran}
\bibliography{mybib}

\end{document}